\documentclass[journal]{IEEEtran}
\usepackage[section] {placeins}
\usepackage{graphicx}
\usepackage{amsmath}
\usepackage{subfigure}
\usepackage{rotating}
\usepackage{multirow}
\usepackage{array}
\usepackage{rotating}
\usepackage{auto-pst-pdf}
\usepackage{graphicx}
\usepackage{float}
\usepackage{pdfpages}

\begin{document}

\title{AMCTD: Adaptive Mobility of Courier nodes in Threshold-optimized DBR Protocol for Underwater Wireless Sensor Networks}

\author{M. R. Jafri$^{1}$, S. Ahmed$^{1,2}$, N. Javaid$^{1,4}$, Z. Ahmad$^{2}$, R. J. Qureshi$^{3}$\\\vspace{0.4cm}
$^{1}$Dept of Electrical Engineering, COMSATS Institute of IT, Islamabad, Pakistan.\\
$^{2}$Abasyn University, Peshawar, Pakistan.\\
$^{3}$SZABIST, Dubai, UAE.\\
$^{4}$CAST, COMSATS Institute of IT, Islamabad, Pakistan.
}

\maketitle

\begin{abstract}
\boldmath
In dense underwater sensor networks (UWSN), the major confronts are high error probability, incessant variation in topology of sensor nodes, and much energy consumption for data transmission. However, there are some remarkable applications of UWSN such as management of seabed and oil reservoirs, exploration of deep sea situation and prevention of aqueous disasters. In order to accomplish these applications, ignorance of the limitations of acoustic communications such as high delay and low bandwidth is not feasible. In this paper, we propose Adaptive mobility of Courier nodes in Threshold-optimized Depth-based routing (AMCTD), exploring the proficient amendments in depth threshold and implementing the optimal weight function to achieve longer network lifetime. We segregate our scheme in 3 major phases of weight updating, depth threshold variation and adaptive mobility of courier nodes. During data forwarding, we provide the framework for alterations in threshold to cope with the sparse condition of network. We ultimately perform detailed simulations to scrutinize the performance of our proposed scheme and its comparison with other two notable routing protocols in term of network lifetime and other essential parameters. The simulations results verify that our scheme performs better than the other techniques and near to optimal in the field of UWSN.
\end{abstract}

\begin{IEEEkeywords}
Weight function, Courier nodes, Underwater Wireless Sensor Networks, Depth-based routing.
\end{IEEEkeywords}

\IEEEpeerreviewmaketitle
\section{Introduction}
\IEEEPARstart{I}{N}
 the recent years, UWSNs have emerged brilliantly for its appliances in management, control and surveillance in selected portions of deep oceans. The dynamic conditions, variations in topologies, energy constraints and high error probability during data forwarding are prominent challenges in the design of routing protocols in UWSN. The acquisition of real-time wireless data access in underwater environment is also a key demand for the future proposals in the field of UWSN routing protocols. Unlike in terrestrial networks, the use of acoustic communication causes a larger propagation delay and low bandwidths, which has to be overcome in the proposed techniques.  Along with the scientific exploration in the deep sea water, oil reserves management and coastline protection are the major demands from UWSN. In the recent literature, distinguished techniques address the problems of high end-to-end delay, multipath fading and other mobility issues. In this paper, we strive to achieve global load balancing utilizing the adaptive movement of courier nodes and implementation of optimal weight computations for the sensor nodes. We also addresses the problems of high transmission and receiving power consumption in UWSN  by employing better coordination among the nodes and minimal data forwarding. In UWSN, recent advances have shown the importance of long-term environment monitoring; demonstrating longer lifetime of network as the key requirement in the future suggestions. The maintenance of higher throughput is also required during the entire network lifetime. DBR ~\cite{1} and EEDBR ~\cite{2} propose the well-organized energy consumption schemes in the stability period but there is a lack of optimality in throughput during the instability period of network. In these techniques, the main load is on the low-depth nodes, causing their rapid energy consumption and coverage holes in network. There is a lack of load balancing in the typical routing protocols due to unequal load distribution among the nodes. Self-management and self-optimization are the major achievements of our proposed technique in changing topology of UWSN. In aqueous environment, the routing protocols are divided into two major categories: localization-free and localization-based routing schemes. Localization-free routing protocols do not assume the localization information of sensor nodes. Our scheme is also categorized in the latter category and its routing is based on the depth information of sensor nodes.
\hfill

\section{Related Work}
Routing protocols play a key role to prolong network lifetime. In this regard, authors in ~\cite{3}, ~\cite{4}, ~\cite{5}, ~\cite{6} and ~\cite{7} proposed different schemes for terrestrial WSNs. In recent researches, it has been proved that the delay-tolerant applications are the major intention of UWSN. Therefore, the notable proposals in underwater routing protocols investigate lack of global load balancing in the network to obtain extended lifetime of network. An eminent technique in localization-free category is Depth-based routing protocol (DBR), based on data forwarding through low-depth sensor nodes. Energy-Efficient depth-based routing scheme (EEDBR) is a constructive framework for maximizing the network lifetime by utilizing both depth and residual energy of the sensor nodes. It minimizes the end-to-end delay along with better energy consumption of the low-depth nodes. Both of the afore-mentioned techniques strive to deal with the challenge of minimizing the load on medium-depth sensor nodes in dense conditions. H2-DAB ~\cite{8} tackle the challenges of UWSN by implementing the dynamic addressing scheme among the sensor nodes without requiring the localization information. Another efficient scheme, R-ERP2R~\cite{9} employs the routing metric based on the physical distances between the nodes and exercises it to accomplish higher throughput in UWSN. It also provides the energy efficient solution for data forwarding along with better link quality. The sensor nodes compute the holding time based on their depth during the optimal forwarder selection to eradicate the needless flooding of data packets. QELAR ~\cite{10} offers an outline for evenly distributed residual energy among the nodes to calculate the reward function. It strives to achieve longer network lifetime by selecting adequate forwarders for source nodes. In dense underwater conditions, PULRP ~\cite{11} provides layered architecture and detailed algorithm to achieve higher throughput among the network. It is free of localization information and do not require fixed routing table, minimizing the overhead in the network. In notable localization based protocols, HH-VBF ~\cite{12} uses the vectors assumptions between the source and the destination nodes, effective for both dense and sparse conditions in the network. It gives a vector-based algorithm to achieve low end-to-end delay in the network in spite of no state information of sensor nodes. Therefore, we propose Adaptive Mobility of Courier nodes in Threshold-optimized Depth-based routing protocol (AMCTD) to attain reduced end-to-end delay, longer stability period and better network lifetime to cope with the varying conditions of UWSNs.

\section{Motivation and Contribution}
In the previous literature of routing protocols in UWSN, there is lot of prominent depth-based routing protocols. However, the efficient energy consumption remains the main acquisition due to necessity of longer network lifetime. DBR and EEDBR struggle to obtain longer network lifetime but the stability period ends quickly due to unnecessary data forwarding and much load on low-depth nodes in UWSN. In the later technique, the nodes with high residual energy expire earlier due to increased load, causing less number of available neighbors for the remaining nodes in sparse condition and immense coverage holes in the core of network. The main deficiency of the depth based routing protocols such as DBR and EEDBR is disorganized instability period due to the quick energy consumption of medium-depth nodes which has been addressed in our proposed scheme. However, EEDBR removes key deficiencies of DBR but it has a lot of imperfections which have been overcome by our scheme as follows:
\begin{itemize}
  \item The adaptive changes in depth threshold removes the problem of lack of availability of threshold based neighbors during the entire network lifetime.
  \item The proficient movement of courier nodes minimizes the end-to-end delay as well as decrements the energy consumption of low-depth nodes utilizing on-spot data collection, making the scheme advantageous for data-sensitive purposes.
  \item The optimal weight computation techniques cause longer stability period and also offer better throughput in the sparse conditions of network.
\end{itemize}
In this paper, we investigate the problems of little stability period, swift energy consumption of low-depth nodes and poor throughput during the instability period caused due to unequal load distribution among the nodes. We promote global load balancing in our proposed scheme by utilizing the optimized movement of courier nodes in sparse conditions. Based on the all-above analysis, this paper presents Adaptive Mobility of Courier nodes in Threshold-optimized Depth-based routing scheme (AMCTD) to accomplish resourceful energy expenditure of nodes in UWSN.

\section{Proposed Routing Protocol: AMCTD}
In this section, we introduces our proposed protocol AMCTD descriptively. in the first section of network initialization, each node computers its weight on the basis of density of network and the movement of courier nodes has been designed. in the second phase of data forwarding, the optimal forwarders are decided on the basis of prioritization of weights of the neighbours of the source node. in the third section of weight updating and depth threshold adaption, network assigns the weights on the prioritization of depth and residual energy is changed according to the sparsity of network. in the last section, the movement of courier nodes and variations in depth threshold of nodes have been devised to cope with the sparsity of network. the afore mentioned steps have been discussed as following.

\subsection{Network Model}
In this section, we introduce our proposed protocol AMCTD descriptively. In the first section of network initialization, each node figures out its weight on the basis of density of network and the courier nodes initiate their schematic movement towards the surface of network. In the second phase of Data forwarding, the source nodes decide the best possible forwarders on the basis of prioritization of weights among their threshold-based neighbors. In the third section of weight updating, network allocates the weights to sensor nodes, based on the prioritization of their depths instead of residual energy to cope with the sparse conditions of network. In the last section of depth threshold adaption, our routing scheme devises the precise course for motorized movement of courier nodes along with variations in depth threshold of nodes to deal with coverage holes created in the later rounds of network lifetime. The afore-mentioned steps have been discussed briefly as following.

\subsection{Network Architecture}
In our first phase, courier nodes devise their schematic sojourn tour in the network as the sensor node broadcast their depth information to the neighbors using hello packets. Therefore, the joint communication between the sensor nodes, courier nodes and the sinks initializes the network operations. The key elements of network architecture design are as follows
\begin{itemize}
  \item Sink sends hello packet to all the nodes to get their vital information.
  \item The network sets down the depth threshold of sensor nodes to 60m to eliminate flooding process.
  \item Each node calculates its weights using the below-mentioned formula and employing the value of Priority value.
\end{itemize}
\begin{equation}
W_{i}=(priority value*R_{i})/(Depth of water-D_{i})
\end{equation}
where $R_{i}$ is the residual energy of node i, $D_{i}$ is the depth of node I and Priority value is a constant . The weight calculation technique minimizes the load on the low-depth nodes, causing the increase in stability period. Moreover, the computational formula for the weight of sensor nodes changes with the varying density of network.
\begin{figure*}[t]
\centering
\includegraphics[height=7cm, width=12cm]{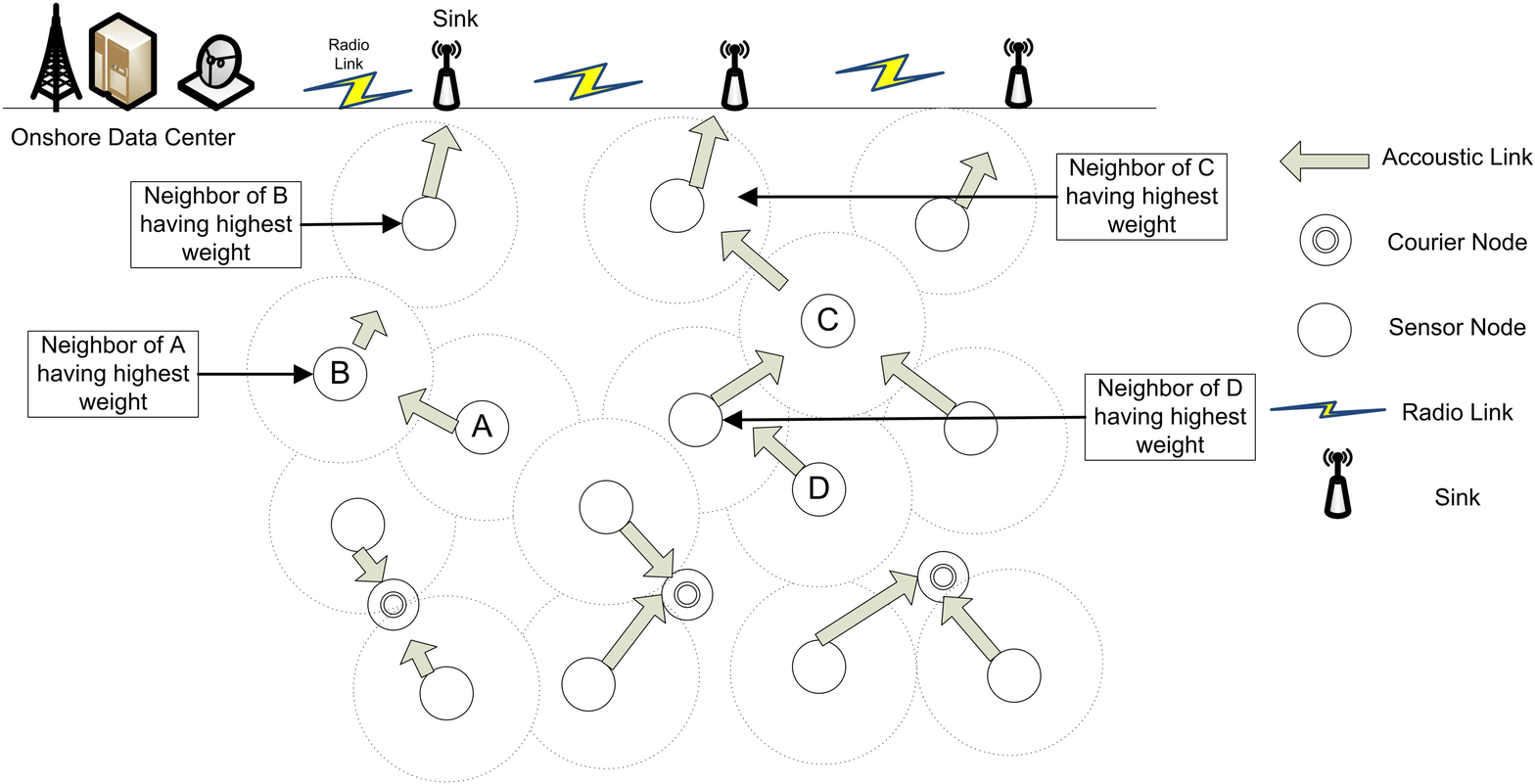}
\caption{Mechanism of Data Transmission in AMCTD}
\end{figure*}

\subsection{Initialization Phase}
During this phase, each sensor node shares its weight and depth information with its neighbors. Employing hello packets transmission, each node identifies its neighbors in transmission range and maintains the separate queue of neighbors under depth threshold to identify the finest forwarder for its data transmission. As the network initializes, the courier nodes start their sojourn movements towards the surface of water. Due to unlimited supply of energy, they aggregate the data of the sensor nodes continuously, transmit it to the sink and then go downward again to start the preceding sojourn tour. At one hand, the courier nodes collect the data and on the other hand change its mobility pattern to adapt with the changing network density. Their schematic mobility model helps out the network to diminish coverage holes creation in the instability period. Each node also finds for the courier node among its neighbors to encourage on-spot data collection.. If courier nodes receive the packet of source node, it transmits acknowledgment to other neighbor nodes to stop further forwarding by any other neighbor of the source node.

\subsection{Network Adaption Specifications and data forwarding}
 After sensing data, node sends its data toward sink using the technique of CSMA/CA. Source node finds the optimal forwarder among its threshold-based neighbors by comparing their weights. The neighbor having the highest weight is elected as forwarder and after receiving the packet it waits for holding time before upward data transmission. It discards the packet on receiving of same packet from any other neighbor node during the holding time duration. The node having lesser weight have much holding time, therefore, the optimal forwarder has smaller holding time duration then the other neighbors of source node. After every 50 rounds, the nodes broadcast hello packet in the network to find the number of dead node by the sink. It is used to cope with the changing conditions of the network and computations of network parameters. Furthermore, if the two neighbors have same depth, the optimal forwarder will be one having more residual energy. If courier node receives the packet, it transmits acknowledgment to other neighbors of source node to eliminate needless forwarding by any other neighbor node. It utilizes the packet ID and source ID to accomplish this purpose causing resourceful energy consumption of the source node. The nodes continue to forward the packet of the source node until it reaches to the  base station or courier node.

\subsection{Weight Updating Phase}
This phase specifies the revisions in weight calculation of sensor nodes according to the altering node density of the network. After the number of dead node increases by 2 \%, each node calculates its weight by the following formula
\begin{equation}
W_{i}=(priority value*D_{i})/R_{i}
\end{equation}
This alteration is used to prioritize the depth among neighbors and to reduce the significance of residual energy in the calculation of the weight. It decreases the load on nodes with high residual energy and low depth to become optimal forwarder for consecutive transmissions. It chiefly plays role in incrementing the instability period of the network.

\subsection{Variation in Depth Threshold and Movement Scheme of Courier Nodes}
To adapt with the coverage problems of network, our design proposes the optimal variation in the depth threshold and the movement pattern of courier nodes, causing augmentation in the instability period. As the number of dead nodes increases by 75 \%, First and third courier node starts to move between the depth of 355m and the bottom of water, collecting the data from the high-depth sensor nodes. It is assumed that these nodes have minimal number of neighbors left alive mostly due to sparsity of the network. The speed of the movement of the courier node is also varied to a higher speed to facilitate the remaining alive nodes. Moreover, the second and fourth courier nodes move vertically between 100m and 200m to accumulate the data from intermediate-depth nodes. The changing movement pattern mainly increases the network lifetime in the later rounds. In this section, we also discuss the well-organized framework for alterations in depth threshold. As the number of dead nodes increases by 2 \%, the depth threshold is decreased to 40m to increase the quantity of threshold-based neighbors of the source nodes. It eases out the forwarding of data in low network density moreover, it is changed to 20 m as the dead nodes increase by 200 in extreme less dense condition to boost the network lifetime. In UWSNs, the network lifetime is of prime importance; hence our scheme proposes the modification in weight calculation again as the number of dead nodes passes 80 \% to prioritize the value of residual energy among the remaining alive neighbors of the source sensor nodes
\begin{equation}
W_{i}=R_{i}/(priority value*D_{i})
\end{equation}
where $R_{i}$ is the residual energy of node i and $D_{i}$ is the depth of node i.

\section{Performance Evaluation and Analysis}
In this section, we assess the performance of AMCTD in underwater sensor networks using MATLAB simulator. We compare the performance of our protocol with other energy efficient depth-based routing protocols such as DBR and EEEDBR. Both of the afore-mentioned protocols are considered to be an imperative landmarks in depth based routing schemes.
\subsection{Simulation Scenarios}
In order to analyze the results in practical scenario, we deploy the network of 225 nodes using random topology in 500mx500m environment. The transmission range of sensor node is 100 meters, following the physical characteristics of underwater acoustic modem. We have adopted the specifications of commercial acoustic modem, LinkQuest UWM1000, where the data generating rate is 1 packet per second. The power consumptions for the transmitting, receiving and idle mode are 2w, 0.1w and 10mw respectively. The initial energy of the node has been set to 70 joules while the packet size is 50 bytes. At the surface of water, the mutual distance between the sinks is 100 meters consecutively. The courier nodes reside in the bottom of water during the network initialization. Then, they start to move upwards, transmit the received data to the sink and then again moves downward toward the bottom. Each node broadcasts hello packets after 50 rounds to check the number of dead nodes from the sink. In every single simulation run, all the nodes of the network sense data and transmit upwards, until it reaches to base station or courier nodes. To ensure efficient media access among the nodes, 802.11-DYNAV ~\cite{13} protocol have been suggested. Each node shares the vital physical metrics, especially depth threshold and weight with its neighbors to keep it informed with the varying circumstances of the network. Our routing algorithm shapes the movement course of courier nodes to compensate with the sparse conditions of the network exploiting the transmission of hello packets. However, there is a trade-off between the overhead (because of Hello packets’ transmission) and the availability of up-to-date information.

The simulation parameters are given in Table 1.
\begin{center}
\begin{tabular}{p{5cm} l  l}
  Table 1&~\\
  Parameters used in Simulations&~\\
  \hline
  Parameter & Value \\
  \hline
  Network size & 500m x 500m \\ 
  Node number & 225 \\ 
  Initial energy of normal nodes & 70J \\ 
  Data aggregation factor & 0.6 \\ 
  Packet size & 50 bytes\\
  Transmission Range & 100 meters\\
  Number of Courier nodes & 4\\
  Number of Simulations & 3\\
  \hline
  ~&~\\
\end{tabular}
\end{center}
\subsection{Performance Metrics}
We demonstrate the following experimental metrics to express the performance of our proposed technique.
\begin{itemize}
  \item Network Lifetime: It is the time duration between the network initialization and the complete energy exhaustion of all the nodes.
  \item Average Energy Consumption: It is the energy consumption of all the active nodes of network in 1 round.
  \item Probability of Dropped packets:  It shows the probability of loss of packets in 1 round.
  \item Number of Dead nodes: It shows the number of dead nodes of the network.
  \item Confidence interval:  It is an interval in which a measurement or trial falls corresponding to a given probability.
\end{itemize}
\subsection{Simulation Results and Analysis}
We compare the network lifetime of our proposed technique AMCTD with DBR and EEDBR. In the simulation of 15000 rounds, nodes have been deployed randomly in every simulated technique. Figure~2 represents the comparison between the network lifetime of AMCTD, EEDBR and DBR. The results of network lifetime comprises on the average of the 3 simulation runs of the afore-mentioned techniques and their comparisons, causing the increase in outcome authenticity. In every single round of simulation, each alive node of the network send the packet, and the packet is forwarded until it reaches to the sink or courier node. In AMCTD, first node dies at about 4500 round meanwhile, it also provides the instability period of about 12000 rounds. After the dying of initial nodes, the adaptive variations in the depth threshold provide longer lifetime to the network. The modifications in weight calculation technique and the prioritization of weight also encourages the efficient instability period. It causes the exclusion of abrupt energy consumption of sensor nodes due to better availability of threshold-based neighbors. In last 3000 rounds, the movement pattern of courier nodes and their sojourn intervals enhance the probability of efficient data transmission. In the previous techniques of DBR and EEDBR, the stability period ends quickly due to prioritization of residual energy or depth solely in the selection of optimal neighbors, which causes inefficient instability period. In our proposed technique, resourceful utilization of energy becomes possible due to modification of depth threshold according to shifting network concentration.
\begin{figure}[!h]
\centering
\includegraphics[height=6.5cm, width=9cm]{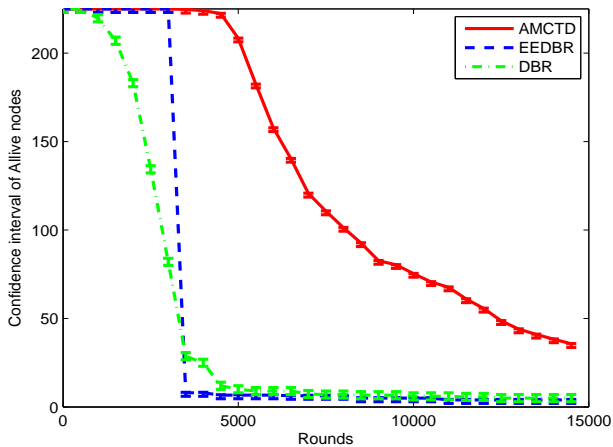}
\caption{Confidence Intervals of number of Alive nodes in AMCTD, EEDBR and DBR}
\end{figure}
In order to show the reliability of the estimates of simulation results, Figure~2 also shows the confidence intervals between network lifetimes of the contrasting techniques.  Confidence intervals consist of a series of values that operate as good estimates of the unknown results of network parameters. The first interval descriptively justifies the larger stability period of our proposed techniques which is better than the other depth-based routing protocols due to efficient energy consumption of the sensor nodes. The intervals throughout the network lifetime illustrate the consistency in the performance of our recommended technique avoiding the sudden collapse of network due to coverage holes creation in the network.
Figure~3 shows the evaluation of dead node variation in AMCTD, EEDBR and DBR alongwith the average results of 3 simulation runs. The implementation of adaptive mobility of courier nodes improves the stability period of AMCTD then that of DBR and EEDBR, which is also caused due to removal of redundant data forwarding by neighbor nodes. The key reasons behind the much capable instability period of our technique then of afore-mentioned schemes are changes in depth threshold and optimal forwarder assortment, prioritizing the sensor node depth in later rounds.
\begin{figure}[!h]
\centering
\includegraphics[height=6.5cm, width=9cm]{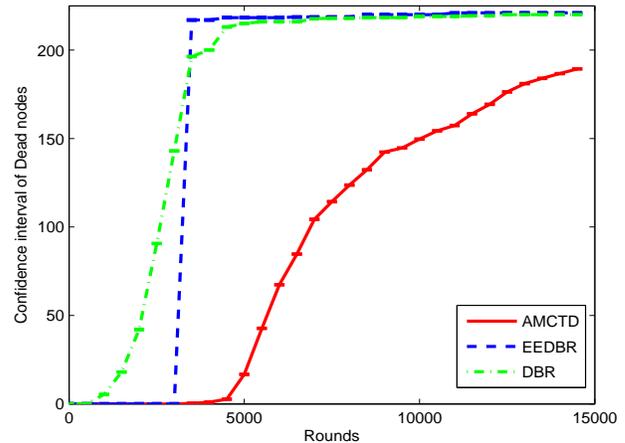}
\caption{Comparison of Dead nodes in AMCTD, EEDBR and DBR}
\end{figure}
The global load balancing is achieved by the low-depth movement of courier nodes, nevertheless in previous schemes low-depth nodes expire earlier due to a large extent of data forwarding. In DBR and EEDBR, high-depth nodes pass away rapidly in later rounds due to minimal quantity of neighbors. It affects the network harshly and the network throughput suddenly reduces because of swift raise in number of dead nodes. The optimized weight computation removes the same node maximal selection as a top forwarder; hence curtailing the load the on nominal nodes and enhancing the network lifetime. The adaptive movement of courier nodes further diminishes the delay in data delivery to the base station. The confidence interval defines the accuracy of results along with removal of sudden collapse of network.
Figure~4 describes the comparison between the average energy consumption of network. In our scheme, energy utilization of sensor nodes is much proficient at start due to effective weight implementation and slighter data forwarding. However, in the previous proposals, uneven energy consumption of nodes is cause due to elevated data forwarding towards the base stations by medium-depth nodes. It sources low stability period along with sharp reduction in the throughput of instability period.
Moreover, the higher energy utilization in our scheme in mid rounds is because of increase in the network throughput. It upholds the throughput and encourages smooth energy consumption by all the source and forwarding nodes. The proficient energy consumption causes enhancement in network lifetime and adjusts with the changing state of network concentration. The average energy consumption also undergoes a sudden drop off in the previous techniques due to the unequal distribution of load on specific low-depth nodes, causing low performance of network. A lot of empty spaces and coverage holes are created in the network during instability period which enlarge the probability for the loss of packets. The confidence intervals confirm that higher energy utilization in the previous techniques during the initial rounds origins the abrupt fall of network performance afterwards along with the poor coverage of network, therefore causing inefficient instability period.
\begin{figure}[!h]
\centering
\includegraphics[height=6.5cm, width=9cm]{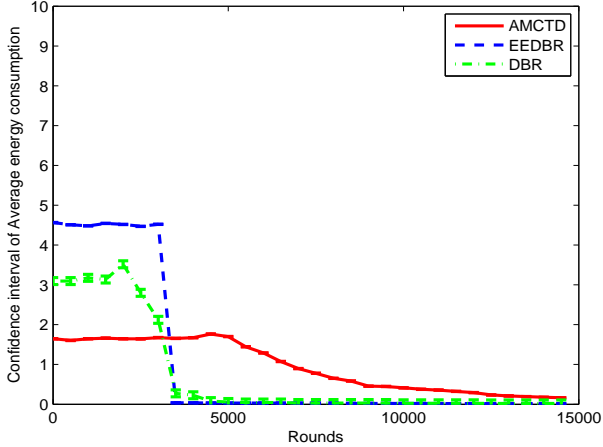}
\caption{Confidence Intervals of Total energy consumption in AMCTD, EEDBR and DBR}
\end{figure}
The confidence intervals of average energy consumption in our scheme encourage the equal energy utilization along the entire lifetime, which minimizes the coverage holes creation and encourages stable instability period.

Figure~5 estimates the throughput of network during the network lifetime. In our proposed technique, the presence of courier nodes and the changes in depth threshold largely enhances the reach ability of packets at base station. Due to less decrement in residual energy of low-depth nodes, the network performance at the later rounds is maintained along with the constant end-to-end delay for packets.
\begin{figure}[!h]
\centering
\includegraphics[height=6.5cm, width=9cm]{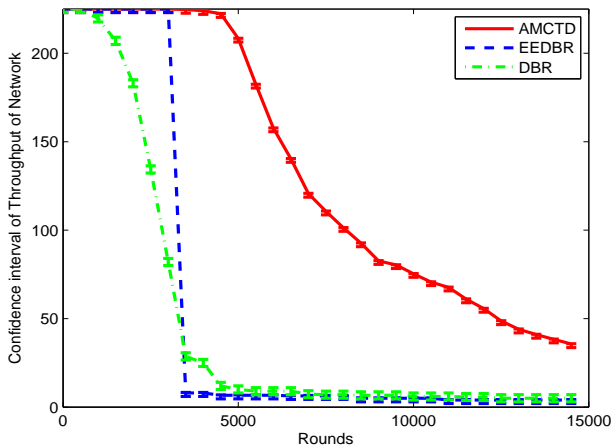}
\caption{Comparison of Network Throughput in AMCTD, EEDBR and DBR}
\end{figure}
The validation improvement due to the computation of average results of 3 simulations not only justifies the advantages of our proposed scheme, but also discriminates its details from the previous depth-based routing protocols.
Figure~6 presents the comparison of Probability of loss of packets in AMCTD, EEDBR and DBR. We have used uniform random model to calculate the packet loss probability in our proposed method. The probability of loss of packets is increased in the previous techniques due to extensive distances between the source and destination nodes along with high depth thresholds.
However, in our proposed framework, the loss probability decreases due to reduced distances involved in data forwarding and less electronic energy cost exploiting the variations in thresholds and mobile courier nodes. Simulation results show that the Probability of loss of packets in AMCTD is better than of EEDBR and DBR of nodes over rounds. The adaptive movement algorithm of courier nodes and the execution of weight function reduce the number of lost packets along with the decrement in end-to-end delay of packets at sink node.
Figure~6 also illustrates the confidence intervals of probability of lost packets of the network. It not only shows the high loss probability in previous techniques due to longer forwarding distances, but also ensures the optimal link judgment in our proposed techniques which increase the network throughput even in the later rounds.
\begin{figure}[!h]
\centering
\includegraphics[height=6.5cm, width=9cm]{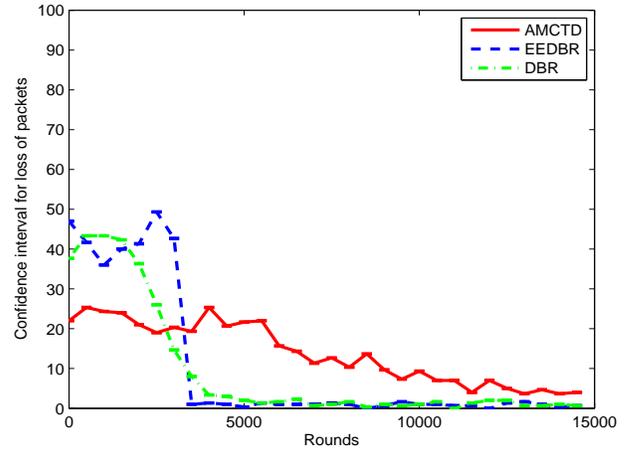}
\caption{Confidence Intervals of Probability of Loss of Packets in AMCTD, EEDBR and DBR}
\end{figure}
\section{Conclusion}
In this paper, we recommend an Adaptive Mobility of Courier nodes in Threshold-optimized Depth-based routing protocol to maximize the network lifetime of UWSN. Our considerations are supportive in decrementing the energy consumption of low-depth sensor nodes specifically in the stability period.  Amendments in depth threshold for the sensor nodes enlarge the number of threshold-based neighbors in the later rounds, hence enhancing the instability period. Optimal weight computation not only provides the global load balancing in the network, but also gives proficient holding-time calculation for the neighbors of source nodes. The adaptive movement of courier nodes upholds the network throughput in the sparse condition of network.
\section{Future Work}
As for future directions, we are striving to design much better courier nodes mobility pattern specifically toward the source nodes in the sparse conditions as well as dense conditions of network for network to perform equally fine in the complete lifetime of UWSNs. We are also planning to integrate MAC protocols ~\cite{14}, ~\cite{15} and ~\cite{16} with our routing scheme in order to facilitate the sensor nodes by the mobility of courier nodes, specifically in the sparse condition to achieve longer network lifetime. Moreover, to increase the network lifetime we are interested to implement routing schemes like ~\cite{17}, ~\cite{18}, ~\cite{19} and ~\cite{20} in UWSNs.

\appendices

\ifCLASSOPTIONcaptionsoff
  \newpage
\fi

%
%

\end{document}